# Two puzzling problems in explanation of the linear Stark effect of hydrogen atom and space quantization of the dipole moment

Pei-Lin You[1]   Xiang-You Huang[2]
1. Institute of Quantum Electronics , Guangdong Ocean University, Zhanjiang 524025, China.
2. Department of Physics, Peking University, Beijing 100871, China.

The linear Stark effect for the first excited state of the hydrogen atom shows that, in the unperturbed states, the atom has a permanent electric dipole moment (EDM) of magnitude $3ea_0$ ($a_0$ is Bohr radius). The EDM is not induce by the external field but is inherent behavior of the atom. But the calculation of quantum mechanics tells us that unperturbed states of hydrogen atom have no EDM! In the effect, two of four states have no energy shift. What are the EDM of the hydrogen atoms in the two states? Quantum mechanics can not answer the problem. The statement that the EDM of the two states is perpendicular to the field only comes from guesses in quantum mechanics. The two problems had puzzled physicists for more than 80 years. By introducing a new parameter this article gives a satisfactory explanation for the effect. Our calculation discovered that, in the unperturbed states, the atoms not only have EDM with $3ea_0$ but also can have only three directions of quantization! It is an entirely unexpected discovery. This research is a vital clue that not all is well with quantum mechanics. We prophesy that hydrogen-like atoms, such as K, Rb and Cs atom etc, may have large EDM of the order of magnitude $ea_0$ in their ground state (see arXiv. 0809.4767, 0810.0770 and 0810.2026).

**1. Introduction**   Quantum mechanics thinks that atoms do not have permanent electric dipole moment (EDM) because of their spherical symmetry. Therefore, there is no polar atom in nature except for polar molecules. A hydrogen atom is the simplest one and physicists understand it very well both in theory and in experiment. However, what problems puzzling us still remain unsolved? Perhaps, a satisfactory interpretation to the linear Stark effect of hydrogen atom is such a problem. The shift in the energy levels of an atom in an electric field is known as the Stark effect. Normally the effect is quadratic in the field strength, but first excited state of the hydrogen atom exhibits an effect that is linear in the strength. This is due to the degeneracy of the excited state.

In the effect the energy levels of hydrogen for weak fields split proportionally to the first power of the field strength (the so-called linear Stark effect). This phenomenon was experimentally shown by Stark from the *Balmer series* in 1913. There was no explanation for the Stark effect in classical theory, only quantum mechanics indicated how to understand this phenomenon. In the effect two of four states have energy level split $\pm 3ea_0\varepsilon$, where $a_0$ is Bohr radius and $\varepsilon$ is the strength of external electric field. This result means that, in the unperturbed state, the hydrogen atom(the quantum number n=2 ) has a EDM whose mean value is $3ea_0=1.59\times 10^{-8}$e.cm[1-4]. On the one hand, this EDM does not depend on the field strength $\varepsilon$, hence it is not induce by the external field but is inherent behavior of the atom[1-4]. L.I. Schiff once stated that "Unperturbed degenerate states of opposite parities, as in the case of the hydrogen atom, can give rise to a permanent electric dipole moment"[1]. L.D. Landay also once stated that "The presence of the linear effect means that, in the unperturbed state, the hydrogen atom has a dipole moment"[3]. On the other hand, the calculation of quantum mechanics tells us that unperturbed degenerate states of hydrogen atom(n=2) have no EDM and has result $\langle\psi_{2lm}|e\mathbf{r}|\psi_{2lm}\rangle=0$, where $\psi_{2lm}$ are four wave functions of unperturbed degenerate states [1, 4]. Due to the EDM of the hydrogen atom is responsible for the presence of linear Stark effect. A hydrogen atom (n=2) with zero EDM how responds to the external field and results in the linear Stark effect ? Quantum mechanics can not answer the problem!

In addition, in the effect, two of four states whose degeneracy was partly removed by the external field, have no energy shift. What are the EDM of the hydrogen atoms in the two states? The usual quantum mechanics can not answer the problem. When Schrödinger established quantum mechanics in 1926, his third paper just explained the linear Stark effect of hydrogen atom (See *Schrödinger, Ann. d. Phys., 80, 437(1926))*. But his explanation contained many guesses. From then on, a large number of brilliant physicists have been making their efforts to the subject but had not succeeded. The two problems had puzzled physicists for more than 80 years.

In the following two sections we shall use the two methods to treat the linear Stark effect of hydrogen atoms. We shall see the difference and the relation between the two methods.

**2. Old method and results** If we neglect spin, in quantum mechanics the wave functions of hydrogen atoms are

$$\Psi(\mathbf{r}, t) = \Psi_{nlm}(r, \theta, \phi) \exp(-i E_n t/h) = R_{nl}(r) Y_{lm}(\theta, \phi) \exp(-i E_n t/h) \tag{1}$$

where $R_{nl}(r)$ is the radial function, $Y_{lm}(\theta, \phi)$ is the spherical harmonics, n is the principal quantum number, $l$ is the angular quantum number, and $m$ is the magnetic quantum number. Since the state of energy level $E_n$ is $n^2$-fold degenerate, so the first excited state (n=2) $\Psi_{2lm}$ is four-fold degenerate. These degenerate wave functions being $\Psi_{200}$, $\Psi_{210}$, $\Psi_{211}$ and $\Psi_{21\text{-}1}$. To solve the problem we use perturbation theory of the degenerate case. When a weak electric field $\varepsilon$ along **Z** axis is applied, the potential energy of the electron in the field, $e\varepsilon z$, is treated as the perturbation H′ = $e\varepsilon z$. The four new approximate wave functions $\Psi_{2(k)}$ are

$$\Psi_{2(1)} = (\Psi_{200} - \Psi_{210})/\sqrt{2}, \quad \Psi_{2(2)} = (\Psi_{200} + \Psi_{210})/\sqrt{2}, \quad \Psi_{2(3)} = \Psi_{211} \text{ and } \Psi_{2(4)} = \Psi_{21\text{-}1} \tag{2}$$

The energy shifts corresponding to the four states are

$$E_{2(1)} = 3ea_o \varepsilon, \quad E_{2(2)} = -3ea_o \varepsilon, \quad E_{2(3)} = E_{2(4)} = 0 \tag{3}$$

In quantum mechanics the explanations of the results are as follows [1-3]. A hydrogen atom in its first excited state (n=2) behaves as though it has a permanent electric dipole moment of magnitude $3ea_o$ that can be oriented in three different ways, where the state $\Psi_{2(1)}$ parallel to the external field, the state $\Psi_{2(2)}$ anti-parallel to the field, and two states $\Psi_{2(3)}$ and $\Psi_{2(4)}$ perpendicular to the field.

On the one hand, the electric dipole moment corresponding to $\Psi_{2(1)}$ and $\Psi_{2(2)}$ are respectively [1-3]

$$\langle \Psi_{2(1)} | e\mathbf{r} | \Psi_{2(1)} \rangle = -3ea_o \mathbf{e}_z \tag{4}$$

$$\langle \Psi_{2(2)} | e\mathbf{r} | \Psi_{2(2)} \rangle = 3ea_o \mathbf{e}_z \tag{5}$$

Which are consistent with the above explanation. On the other hand, the electric dipole moment corresponding to $\Psi_{2(3)}$ and $\Psi_{2(4)}$ are respectively

$$\langle \Psi_{2(3)} | e\mathbf{r} | \Psi_{2(3)} \rangle = \langle \Psi_{2(4)} | e\mathbf{r} | \Psi_{2(4)} \rangle = 0 \tag{6}$$

The result shows that the hydrogen atom of the two states has no electric moment. The result is inconsistent with the above explanation. Furthermore, for any linear combinations of $\Psi_{2(3)}$ and $\Psi_{2(4)}$ we still have

$$\langle C_3 \Psi_{2(3)} + C_4 \Psi_{2(4)} | e\mathbf{r} | C_3 \Psi_{2(3)} + C_4 \Psi_{2(4)} \rangle = 0 \tag{7}$$

where $C_3$ and $C_4$ are normalization constants. The result is also inconsistent with the above explanation. No one will give you any deeper explain of this effect. No one has found any more basic mechanism from which these results can be deduced! The above saying that the EDM of the two states $\Psi_{2(3)}$ and $\Psi_{2(4)}$ perpendicular to the field only comes from guesses in quantum mechanics.

**3. New method and results**   We consider a Hermitian operator *L*. In quantum mechanics the mean value <*L*> of the operator is defined as

$$\langle L \rangle = \int \psi^* L \psi \, dv \tag{8}$$

Because of the electric dipole moment $\langle er(t) \rangle = \langle \psi_{2lm} | er | \psi_{2lm} \rangle = 0$, we have to determine EDM of an atom in a different manner, namely allow the special linear combinations of the degenerate states to replace $\psi_{2lm}$ and $\psi_{2lm}^*$ there. In our method, $\mathbf{d} = -e\mathbf{r}$ is the dipole moment operator of an atom, $-e$ is the charge of the electron, and $\mathbf{r}$ is the position vector of the electron relative to the nucleus[4]. The instantaneous electric dipole moment of an atom is

$$\mathbf{d}(t) = -\text{Re} \int \Phi^*(\mathbf{r},t) \, e\mathbf{r} \, \Psi(\mathbf{r},t) \, dv \tag{9}$$

where the sign Re means the real part of the integral, $\Psi(\mathbf{r},t)$ and $\Phi(\mathbf{r},t)$ are described by the following two wave functions respectively

$$\Psi(\mathbf{r},t) = \Psi_{nlm}(r, \theta, \phi) \exp[im\phi_0] \exp[-iE_n t/h] \tag{10}$$

$$\Phi(\mathbf{r},t) = \sum_{l'=0}^{n-1} \sum_{m'=0}^{\pm l'} \Psi_{nl'm'}(r, \theta, \phi) \exp(im'\phi_0) \exp[-iE_n t/h] \tag{11}$$

Where $\phi_0$ is new parameter introduced in the wave function, m and m' are the magnetic quantum number. The parameter $\phi_0$ of every atom is to be treated as distinct from every other one. Due to the parameter $\phi_0$ appears in phase factors and does not damage the unlocalization of the wave functions. When the parameter is averaged, the new theory will transit to the usual quantum mechanics. Writing the permanent electric dipole moment of an atom as $\langle \mathbf{d} \rangle$, it is given by averaging $\mathbf{d}(t)$ over time t. So the EDM of a hydrogen atom of the unperturbed states $\Psi_{nlm}$ is

$$\langle \mathbf{d} \rangle = -\text{Re} \int \sum_{l'=0}^{n-1} \sum_{m'=0}^{\pm l'} \Psi_{nl'm'}^*(r, \theta, \phi) \exp(-im'\phi_0) e\mathbf{r} \Psi_{nlm}(r, \theta, \phi) \exp(im\phi_0) \, dv \tag{12}$$

When n=2, Eq.(12) becomes

$$\langle \mathbf{d} \rangle = -\text{Re} \int (\Psi_{211}^* e^{-i\phi_0} + \Psi_{21-1}^* e^{i\phi_0} + \Psi_{210}^* + \Psi_{200}^*) e\mathbf{r} \, \Psi_{2lm} \, e^{im\phi_0} \, dv \tag{13}$$

We obtain the position vector for the spherical coordinates $(r, \theta, \phi)$:

$$\mathbf{r} = r \sin\theta \cos\phi \, \mathbf{e}_x + r \sin\theta \sin\phi \, \mathbf{e}_y + r \cos\theta \, \mathbf{e}_z \tag{14}$$

In order to make calculation easy, we let[5]

$$\mathbf{e}_1 = (-\mathbf{e}_x + i\mathbf{e}_y)/\sqrt{2}, \quad \mathbf{e}_{-1} = (\mathbf{e}_x + i\mathbf{e}_y)/\sqrt{2} \tag{15}$$

and have the relation[5]

$$\mathbf{r} = r\sqrt{4\pi/3} \, (Y_{11}(\theta, \phi) \mathbf{e}_1 + Y_{1-1}(\theta, \phi) \mathbf{e}_{-1} + Y_{10}(\theta, \phi) \mathbf{e}_z) \tag{16}$$

Substituting this value into Eq.(13), the EDM of a hydrogen atom in the unperturbed state $\Psi_{2lm}$ are respectively

$$\langle \mathbf{d} \rangle = 3ea_o \mathbf{e}_z \quad \text{for} \quad \Psi_{200} \tag{17}$$

$$\langle \mathbf{d} \rangle = 3ea_o \mathbf{e}_z \quad \text{for} \quad \Psi_{210} \tag{18}$$

$$\langle \mathbf{d} \rangle = -3ea_o (\cos\phi_0 \, \mathbf{e}_x - \sin\phi_0 \, \mathbf{e}_y)/\sqrt{2} \quad \text{for} \quad \Psi_{211} \tag{19}$$

$$\langle \mathbf{d} \rangle = 3ea_o (\cos\phi_0 \, \mathbf{e}_x - \sin\phi_0 \, \mathbf{e}_y)/\sqrt{2} \quad \text{for} \quad \Psi_{21-1} \tag{20}$$

Table 1 gives old and new calculation results of EDM of four unperturbed states for hydrogen atom (n=2).

**Table 1 Old and new results of EDM of four unperturbed states for hydrogen atom(n=2)**

| State | Old result | New result |
|---|---|---|
| $\Psi_{200}$ | 0 | $3ea_o \mathbf{e}_z$ |
| $\Psi_{210}$ | 0 | $3ea_o \mathbf{e}_z$ |
| $\Psi_{211}$ | 0 | $-3ea_o (\cos\phi_0 \, \mathbf{e}_x - \sin\phi_0 \, \mathbf{e}_y)/\sqrt{2}$ |
| $\Psi_{21-1}$ | 0 | $3ea_o (\cos\phi_0 \, \mathbf{e}_x - \sin\phi_0 \, \mathbf{e}_y)/\sqrt{2}$ |

The result shows that a hydrogen atom of the first excited state, in the unperturbed state, has the large EDM.

When the electric field is applied, two approximation wave functions are

$$\Psi_{n(k)}(\mathbf{r},t) = \Psi_{n(k)}(r, \theta, \phi) \exp(im'\phi_0) \exp[-i E_n t/h] \qquad (21)$$

$$\Phi(\mathbf{r},t) = \sum_{k'=1}^{n^2} \Psi_{n(k')}(r, \theta, \phi) \exp(im'\phi_0) \exp[-i E_n t/h] \qquad (22)$$

where $k'$ is the quantum number, $m'$ is the magnetic quantum number, $\phi_0$ still is new parameter introduced in the theory. The EDM of the hydrogen atom corresponding to the perturbed states $\Psi_{n(k)}$ are

$$\langle \mathbf{d} \rangle = -\mathrm{Re} \int \sum_{k'=1}^{n^2} \Psi^*_{n(k')}(r, \theta, \phi) \exp(-im'\phi_0) e\mathbf{r} \, \Psi_{n(k)}(r, \theta, \phi) \exp(im\phi_0) \, dv \qquad (23)$$

When n=2, Eq.(23) becomes

$$\langle \mathbf{d} \rangle = -\mathrm{Re} \int (\Psi^*_{2(1)} + \Psi^*_{2(2)} + \Psi^*_{2(3)} e^{-i\phi_0} + \Psi^*_{2(4)} e^{i\phi_0}) e\mathbf{r} \, \Psi_{2(k)} e^{im\phi_0} \, dv \qquad (24)$$

From Eq.(24) the four EDM corresponding to the perturbed states $\Psi_{2(k)}$ are respectively

$$\langle \mathbf{d} \rangle = -3ea_0 \mathbf{e}_z \quad \text{for } \Psi_{2(1)}, \qquad \langle \mathbf{d} \rangle = -3ea_0(\cos\phi_0 \mathbf{e}_x - \sin\phi_0 \mathbf{e}_y) \quad \text{for } \Psi_{2(3)} \qquad (25)$$

$$\langle \mathbf{d} \rangle = 3ea_0 \mathbf{e}_z \quad \text{for } \Psi_{2(2)}, \qquad \langle \mathbf{d} \rangle = 3ea_0(\cos\phi_0 \mathbf{e}_x - \sin\phi_0 \mathbf{e}_y) \quad \text{for } \Psi_{2(4)} \qquad (26)$$

Table 2 gives old and new results of four perturbed states for hydrogen atom(n=2) when the field is applied.

**Table 1  Old and new results of EDM of four perturbed states for hydrogen atom(n=2)**

| State | Old result | New result |
| --- | --- | --- |
| $\Psi_{2(1)}$ | $-3ea_0 \mathbf{e}_z$ | $-3ea_0 \mathbf{e}_z$ |
| $\Psi_{2(2)}$ | $3ea_0 \mathbf{e}_z$ | $3ea_0 \mathbf{e}_z$ |
| $\Psi_{2(3)}$ | 0 | $-3ea_0(\cos\phi_0 \mathbf{e}_x - \sin\phi_0 \mathbf{e}_y)$ |
| $\Psi_{2(4)}$ | 0 | $3ea_0(\cos\phi_0 \mathbf{e}_x - \sin\phi_0 \mathbf{e}_y)$ |

## 4. Discussion

① Before the electric field is applied, the hydrogen atom of four unperturbed degenerate states has large EDM with $3ea_0$. These EDM can respond to the external electric field and result in the linear Stark effect. **They can be oriented only three directions of quantization, this is called space quantization of the dipole moment, namely the angle θ between the dipole moment vector and Z axis can have only certain values:**

$$\sin\theta = m/\sqrt{l(l+1)} \qquad (27)$$

$$\theta = 0 \text{ for } \Psi_{200} \text{ and } \Psi_{210} \qquad (28)$$

$$\theta = \pi/4 \text{ or } 3\pi/4 \text{ for } \Psi_{211} \qquad (29)$$

For $\Psi_{21-1}$ θ =5π/4 when θ =π/4 for $\Psi_{211}$ and θ =7π/4 when θ =3π/4 for $\Psi_{211}$ (30)

Notice that in the state $\Psi_{200}$ the magnetic quantum number $m=0$ is rigorously but the angular quantum number $l=0$ is approximately. Indeed we can infer from (19) and (20) the x-y plane projection of a vector dipole moment with absolute value $3ea_0$. These results can be interpreted in Eq.(27). Different hydrogen atoms with same state $\Psi_{211}$ or $\Psi_{211}$ distribute on a cone around the Z axis. **These results exceeded all physicist's expectation and it is an entirely unexpected discovery.**

② When the weak electric field is applied, all atoms of states $\Psi_{2(k)}$ have the EDM with $3ea_0$. The EDM of atom in state $\Psi_{2(1)}$ parallel to the field, the EDM in state $\Psi_{2(2)}$ anti-parallel to the field and the EDM in state $\Psi_{2(3)}$ and $\Psi_{2(4)}$ are perpendicular to the field. **Eq.(25) and (26) give us an even clearer understanding for the linear Stark effect of hydrogen atom.**

③ Different hydrogen atoms have different parameter $\phi_0$. **From Eq.(25) and Eq.(26) we see that when the parameter $\phi_0$ was averaged, our methods will gained the same results as usual quantum mechanics.** Because of our method allows new parameters to be introduced in it and contains more information of the atom than the usual quantum mechanics. As early as 1926, Schrödinger explained the linear Stark effect of hydrogen atom but his explanation contained many guesses. From then on, there has never been a satisfactory explanation for the effect. On this point our calculation result are both spectacular and convincing. **This research is a vital clue that not all is well with quantum mechanics. It will open a new window for us to view atoms and quantum mechanics again.**

④In order for an atom or elementary particle to possess a permanent electric dipole moment (EDM), time reversal (T) symmetry must be violated, and through the CPT theorem CP(charge conjugation and parity) must be violated as well[6]. The currently accepted Standard Model of Particle Physics predicts unobservable the dipole moments of an atom, therefore, EDM experiments are an ideal probe for new physics beyond the Standard Model. Experimental searches for an EDM of diamagnetic atoms, the most sensitive of which is done with $^{199}$Hg(the result is d(Hg)=-[1.06±0.49 (stat)±0.40 (syst)]×$10^{-28}$e.cm )[6,7]. Experiments to search for an EDM of atom began many decades ago, no large EDM has yet been found [6-9].The radius of the hydrogen atom of the first excited state is $r_H$ = 4$a_o$ =2.12×$10^{-8}$ cm, it is almost the same as the radius of $^{199}$Hg ($r_{Hg}$ =1.51×$10^{-8}$ cm)[10], but the discrepancy between their EDM is by some twenty orders of magnitude! How do explain this inconceivable discrepancy? The current theory can not answer the problem! The current theory thinks that in quantum mechanics there is no such concept as the path of an electron[3]. However, a hydrogen atom (n=2) has a nonzero EDM in the semi-classical theory of atom. The electron in a hydrogen atom (n=2) moves along a quantization elliptic orbit. We can draw a straight line perpendicular to the major axis of the elliptic orbit through the nucleus in the orbital plane. The straight line divides the elliptic orbit into two parts that are different in size. According the law of conservation of angular momentum, the average distance between the moving electron and the static nucleus is larger and the electron remains in the large part longer than in the small part. As a result, the time-averaged value of the electric dipole moment over a period is nonzero for the atom.

⑤Because of $<\psi_{2lm}|er|\psi_{2lm}>$=0 only means that the average EDM of a large number of hydrogen atoms is zero, but does not mean that the EDM of a single hydrogen atom certainly is zero. Perhaps a hydrogen-like atom may have large EDM. The alkali atoms having only one valence electron in the outermost shell can be described as hydrogen-like atom[2]. Since the quantum number of the ground state alkali atoms are n≥2 rather than n=1(this is 2 for Li, 3 for Na, 4 for K, 5 for Rb and 6 for Cs), as the excited state of the hydrogen atom. In the Sommerfeld picture the valence electron moves along a highly elliptical orbit, the so-called diving orbits, approach the nucleus, as the excited state of the hydrogen atom. So we prophesy that ground-state neutral alkali atoms, such as K, Rb and Cs atoms etc, may have large EDM of the order of magnitude $ea_o$[ 11]. If the prophecy is true, the EDM of these atoms can be examined by measuring capacitance at low frequency [12].Experiments to search for the EDM of the ground state K, Rb and Cs atoms have been carried out. Similar results have been obtained [13-17]. Our experimental results gave clear evidence for CP (charge conjugation and parity) violation in K, Rb and Cs atoms. Few experiments in atomic physics have produced results as surprising as there. T**his result is the product of eight years of intense research and is a classic example of how understanding of the matter-antimatter imbalance in the Universe through atomic physics experiments**[13-15].

**Acnowledgement**     The authors thank to our colleagues Zhao Tang , Rui-Hua Zhou, Zhen-Hua Guo, Ming- jun Zheng, Xue-ming Yi, , Xing Huang, and Engineer Jia You for their help in the work.